\documentclass[aps,pra,twocolumn,superscriptaddress,showpacs,nofootinbib]{revtex4-1}

\usepackage{graphicx,lipsum}
\usepackage{subfigure}
\usepackage{amsmath,color}
\def\gappeq{\mathrel{ \rlap{\raise.5ex\hbox{$>$}}
                      {\lower.5ex\hbox{$\sim$}} } }
\def\lappeq{\mathrel{ \rlap{\raise.5ex\hbox{$<$}}
                      {\lower.5ex\hbox{$\sim$}} } }

\usepackage{color}

\newcommand{\del}[1]{\textcolor{red}{}}

\begin{document}

\title{Controllable non-local interactions between dark solitons in dipolar condensates}

\author{T. Bland} 
\affiliation{Joint Quantum Centre Durham--Newcastle, School of Mathematics and Statistics, Newcastle University, Newcastle upon Tyne, NE1 7RU, United Kingdom}
\author{M. J. Edmonds}
\affiliation{Joint Quantum Centre Durham--Newcastle, School of Mathematics and Statistics, Newcastle University, Newcastle upon Tyne, NE1 7RU, United Kingdom}
\author{N. P. Proukakis}
\affiliation{Joint Quantum Centre Durham--Newcastle, School of Mathematics and Statistics, Newcastle University, Newcastle upon Tyne, NE1 7RU, United Kingdom}
\author{A. M. Martin}
\affiliation{School of Physics, University of Melbourne, Victoria 3010, Australia}
\author{D. H. J. O'Dell}
\affiliation{Department of Physics and Astronomy, McMaster University, Hamilton, Ontario, L8S 4M1, Canada}
\author{N. G. Parker}
\affiliation{Joint Quantum Centre Durham--Newcastle, School of Mathematics and Statistics, Newcastle University, Newcastle upon Tyne, NE1 7RU, United Kingdom}

\pacs{03.75.-b,03.75.Lm,05.45.Yv}

\begin{abstract}
We study the family of static and moving dark solitons in quasi-one-dimensional dipolar Bose-Einstein condensates, exploring their modified form and interactions.  The density dip of the soliton acts as a giant anti-dipole which adds a non-local contribution to the conventional local soliton-soliton interaction.  We map out the stability diagram as a function of the strength and polarization direction of the atomic dipoles, identifying both roton and phonon instabilities.  Away from these instabilities, the solitons collide elastically.  Varying the polarization direction relative to the condensate axis enables tuning of this non-local interaction between repulsive and attractive; the latter case supports unusual dark soliton bound states.  Remarkably, these bound states are themselves shown to behave like solitons, emerging unscathed from collisions with each other.

\end{abstract}

\maketitle

\section{Introduction}

Solitons - waves which propagate without dispersion - are a paradigm of nonlinear physics, occurring across systems as varied as water, optical fibres, spin chains, the human circulatory system and atomic Bose-Einstein condensates (BECs) \cite{dauxois_2006}.  Much of the soliton behaviour across these diverse systems, including their interactions and collisions, are universal \cite{stegeman_1999}.  The extreme controllability of the nonlinearity, dimensionality and external potential in atomic BECs makes them an ideal playground for studying solitons \cite{frantzeskakis_2008,abdullaev_2005,frantzeskakis_2010,billam_2013}, with experimental demonstrations of bright \cite{strecker_2002,khaykovich_2002,eiermann_2004,cornish_2006, marchant_2013,mcdonald_2014}, dark \cite{burger_1999,denschlag_2000,becker_2008,stellmer_2008,weller_2008} and dark-bright \cite{becker_2008,hamner_2011} solitons.  Moreover, these ``matter wave solitons" provide insight into the interplay of solitons with quantum coherence \cite{anglin_2008}, and have potential applications in interferometry \cite{strecker_2002,negretti_2004,mcdonald_2014,helm_2015}, surface force detection \cite{cornish_2008} and as robust quantum-information carriers \cite{lewenstein_2009}.

In conventional  media with local nonlinearity, soliton behaviour is well-established \cite{dauxois_2006,drazin_1989}, e.g., solitons interact only at short distances (when their field profiles overlap).  A current direction in soliton research is the addition of non-local nonlinearity.  This enriches the soliton behaviour, e.g., promoting non-local interactions between solitons, and offers prospects for studying wave analogs of particles with long-range interactions and emulating complex nonlinear networks \cite{rotschild_2006}. To date, such ``non-local solitons'' have been observed in liquid crystals \cite{peccianti_2002,cao_2009,piccardi_2011}, thermo-nonlinear optical fibres \cite{rotschild_2006} and liquids \cite{dreischuh_2006}, and optical ring fibres \cite{jang_2013}.  In these cases, the non-local response is provided by molecular re-orientations, heat conduction and acoustic waves, respectively.  Typically these nonlinearities lead to retarded interactions.

The advent of BECs of atoms with sizeable magnetic dipole moments - $^{52}$Cr \cite{griesmaier_2005,beaufils_2008}, $^{164}$Dy \cite{lu_2011,kadau_2015} and $^{168}$Er \cite{aikawa_2012} - allows for the study of superfluids with non-local nonlinearities \cite{lahaye_2009}, arising from the dipole-dipole (DD) atomic interactions. Remarkably, the ratio of local to non-local interactions can be directly tuned through Feshbach resonances \cite{koch_2008}.   This has opened the door to experimental observations of magnetostriction \cite{stuhler_2005}, anisotropic collapse \cite{lahaye_2008} and self-organized droplet phases \cite{kadau_2015}.  A host of predictions such as 2D bright solitons \cite{pedri_2005} and stabilised 3D dark solitary waves \cite{nath_2008} await experimental verification.

In this paper we obtain numerically the family of static and moving dark solitons in a quasi-1D dipolar BEC as stationary solutions in static and moving frames.  We map out the stablity diagram and form of the soliton solutions as a function of the strength and polarization direction of the dipoles, covering experimentally-relevant parameters.   The dark solitons can acquire dramatically modified profiles, including ripples close to the roton instability.  Moreoever we show analytically that the absence of atoms in a dark soliton causes it to act like a giant anti-dipole \cite{klawunn_2008}.  This induces a non-local contribution to the soliton-soliton interactions which scales as $1/z^3$.  This non-local contribution is effectively instantaneous and experimentally tunable in both strength and sign.  When the non-local interaction between the solitons is attractive, the balance with the conventional repulsive soliton-soliton interaction supports bound states of two dark solitons, for which no analog exists in conventional condensates.  We show that these bound states themselves have soliton-like properties in that they emerge unscathed from collisions with each other.

\section{Theoretical framework}
We consider a dilute, ultracold BEC of atoms with mass $m$ and polarized magnetic dipoles.  At such low energies the interactions between the atoms can be described using a universal pseudo-potential \cite{lahaye_2009},
\begin{equation}
U({\bf r}-{\bf r}') = g\delta({\bf r}-{\bf r}')+U_{\rm dd}({\bf r}-{\bf r}').  
\end{equation}
The first term is a contact interaction accounting for the van der Waals interactions characterized by coefficient $g$.  This gives rise to a local nonlinearity/mean-field potential $g n({\bf r})$, where $n({\bf r})$ is the atomic density.  The second term gives the  dipole-dipole interactions, with $U_{\rm dd}({\bf r}-{\bf r}')=C_{\rm dd}(1-3 \cos^2\theta)/4\pi|{\bf r}-{\bf r}'|^3$, where $\theta$ is the angle between the polarization direction and the inter-atom vector ${\bf r}-{\bf r}'$, and $C_{\rm dd}$ (which is conventionally positive) characterizes the strength of the dipoles.  This contribution to the interactions gives rise to a non-local nonlinearity/dipolar mean-field potential $\Phi({\bf r})=\int U_{\rm dd}({\bf r}-{\bf r}')n({\bf r}')~{\rm d}{\bf r}' $.  At the magic angle $\theta_{\rm m}\approx 53^o$ the dipole-dipole interactions disappear.  For $\theta<\theta_{\rm m}$ the dipoles lie dominantly head-to-tail and attract, while for $\theta>\theta_{\rm m}$ they lie dominantly side-by-side and repel.  However, by fast rotation of the polarization direction the time-averaged dipole-dipole interaction can be effectively reversed ($C_{\rm dd}<0$), in which case the dipoles repel when head-to-tail and attract when side-by-side \cite{giovanazzi_2002}. 

In three-dimensional geometries, dark solitons are prone to transverse excitation of the nodal line, the so-called ``snake instability''.  However, in quasi-one-dimensional geometries this instability is prevented and dark solitons become long-lived  \cite{frantzeskakis_2010,weller_2008,muryshev_2002}.   We consider such a quasi-1D waveguide geometry, aligned along $z$.  The confinement in the transverse directions is assumed to be harmonic with the form, $V({\bf r})= m\omega_\perp^2 r^2/2$, where $\omega_\perp$ is the corresponding trap frequency.  The quasi-1D limit is reached when this confinement is sufficiently strong  ($\hbar \omega_\perp \gg \mu$, where $\mu$ is the BEC chemical potential) that the condensate wavefunction approaches the  harmonic oscillator state in the transverse direction \cite{gorlitz_2001,parker_2008}, with lengthscale $l_\perp=\sqrt{\hbar/m \omega_\perp}$.    In this regime the 3D condensate can be parameterized via a 1D mean-field wavefunction $\psi(z,t)$ which obeys an effective 1D Gross-Pitaevskii equation (GPE) \cite{giovanazzi_2004,sinha_2007},
\begin{equation}
i\hbar \partial_t \psi=\left(-\dfrac{\hbar^2}{2m}\partial_{zz}^2+\frac{g}{2 \pi l_{\perp}^2}|\psi|^2 + \Phi_{\rm 1D} \right)\psi.
\label{eqn:gpe}
\end{equation}
Here $\Phi_{\rm 1D}(z,t)=\int U_{\rm 1D}(z-z') |\psi|^2~{\rm d}z'$ is the effective 1D dipolar potential, with associated pseudo-potential \cite{deuretzbacher_2010},
\begin{equation}
U_{\text{1D}}(u){=}U_0\left[2u{-}\sqrt{2\pi}(1{+}u^{2})e^{u^{2}/2}\text{erfc}\left(\frac{u}{\sqrt{2}}\right)+\frac{8}{3}\delta(u)\right], 
\label{eqn:U}
\end{equation}
where $u=|z|/l_\perp$ and $U_0=C_{\rm dd}(1+3\cos 2\theta)/{32\pi l_{\perp}^{3}}$.  While the above mean-field model ignores finite temperature effects, it is worth noting that the salient physical behaviour of the (non-dipolar) dark solitons observed experimentally to date are well-described by the zero-temperature mean-field Gross-Pitaevskii model \cite{burger_1999,denschlag_2000,becker_2008,stellmer_2008,weller_2008}.

\section{Stability of the homogeneous system}

The stability of the homogeneous 1D dipolar BEC has been established in Refs. \cite{giovanazzi_2004,sinha_2007} for dipoles aligned along $z$ ($\theta=0$); here we map out the full parameter space ($0 \leq \theta \leq \pi$).  The ground state has uniform 1D density $n_0$.   Its chemical potential (the eigenvalue associated with the right-hand side of Eq. (\ref{eqn:gpe})) is $\mu_0 = n_0 g /2\pi l_\perp^2+\Phi_0$, where the first term represents the van der Waals mean-field potential and $\Phi_0=-C_{\rm dd}n_0 [1+3 \cos 2 \theta]/24 \pi l_\perp^2$ represents the dipole-dipole mean-field potential. The characteristic length and speed scales are the healing length $\xi=\hbar/\sqrt{m \mu_0}$ and speed of sound $c=\sqrt{\mu_0 /m}$; a time-scale follows as $\tau=\xi /c$.  We parameterise the transverse BEC size via $\sigma=l_\perp/\xi$ where the quasi-1D approximation requires $\sigma\lappeq1$ \cite{gorlitz_2001}.   We take the arbitrary value $\sigma=0.2$ throughout this paper, but our qualitative findings are independent of $\sigma$.

It is convenient to parameterize the dipole strength via the ratio $\varepsilon_{\rm dd} = C_{\rm dd}/3g$ \cite{lahaye_2009}.  In the dipolar condensates produced to date, this ratio has the natural value $\varepsilon_{\rm dd}=0.16$ for $^{52}$Cr \cite{griesmaier_2006}, $\varepsilon_{\rm dd}=0.4$ for $^{168}$Er \cite{aikawa_2012} and $\varepsilon_{\rm dd}=1.4$ for $^{164}$Dy\cite{lu_2011,tang_2015}.  However, $g$ and $C_{\rm dd}$ can both be tuned in size and sign \cite{lahaye_2009,giovanazzi_2002}, and so we will typically consider a range of $\varepsilon_{\rm dd}$ with both negative and positive $g$ and $C_{\rm dd}$  (later, when considering soliton collisions and bound states, we will focus on the parameters for $^{168}$Er, and comment on the wider dependence on $\varepsilon_{\rm dd}$).   

\begin{figure}[t]
\centering
\includegraphics[width=1\columnwidth]{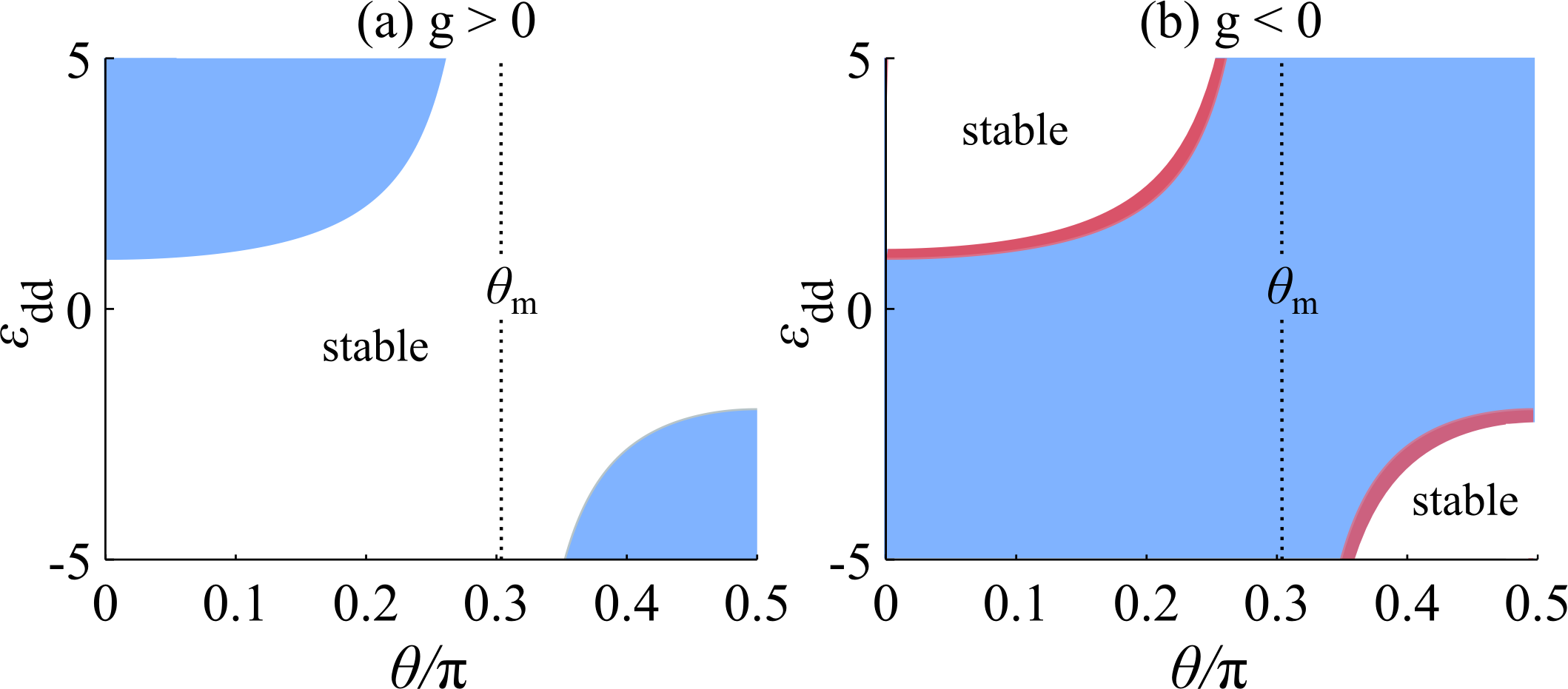}
\caption{(Color online) Stability diagrams in the $(\theta,\varepsilon_{\rm dd})-$plane for the homogeneous quasi-1D dipolar BEC for (a) $g>0$ and (b) $g<0$.  Depicted are the regions of stability (white), phonon instability (blue) and roton instability (red).  The magic angle $\theta_{\rm m}$ is highlighted.}
\label{fig:stab}
\end{figure}

The homogeneous condensate is unstable for certain $C_{\rm dd}$, $g$ and $\theta$ \cite{giovanazzi_2004,sinha_2007}; the stability in the $(\theta,\varepsilon_{\rm dd})$-plane is depicted in Fig.~\ref{fig:stab} for (a) $g>0$ and (b) $g<0$.  The condensate suffers two key instabilities - the phonon instability and the roton instability.  The phonon instability refers to unstable growth of low $k$-modes.  It arises when the net interactions become attractive, i.e. when $\mu_0<0$ (blue shaded regions).  Consider, for example, conventional dipoles ($C_{\rm dd}>0$): for $\theta=0$, the phonon instability arises when the attraction of the end-to-end dipoles dominates the van der Waals repulsion, while for $\theta=\pi/2$ it arises when the repulsive side-by-side dipoles are outweighed by attractive van der Waals interactions.  The reverse is true for $C_{\rm dd}<0$.  In both cases a long wavelength collapse is induced.  

\begin{figure*}
\centering
\includegraphics[width=1.6\columnwidth]{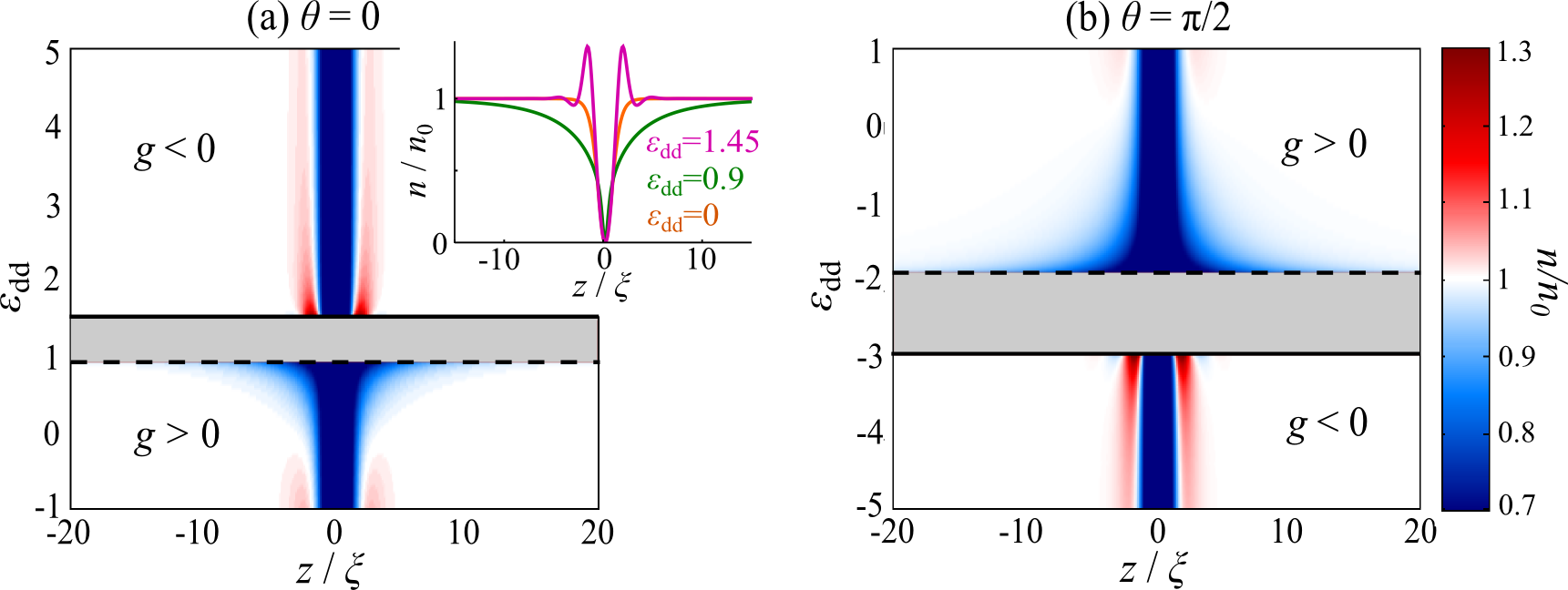}
\caption{(Color online) Density profiles $n(z)$ of the $v=0$ soliton for (a) $\theta=0$ and (b) $\theta=\pi/2$, as a function of $\varepsilon_{\rm dd}$.  The band (grey) of instability is bounded by the onset of the phonon and roton instabilities (dashed and solid lines, respectively).  Inset: density profile for no dipoles ($\varepsilon_{\rm dd}=0$) and close to the instabilities ($\varepsilon_{\rm dd}=0.9, 1.45$).  Note that the unit of length $\xi$ is itself a function of $\varepsilon_{\rm dd}$ and $\theta$.  
}\label{fig:1sol}
\end{figure*}

The dispersion relation can also feature a roton-like dip at finite momenta \cite{giovanazzi_2004,sinha_2007} which, for certain parameter regimes (red regions in Fig.~\ref{fig:stab}(b)) can touch zero energy, signalling the unstable growth of finite momentum modes, i.e. the roton instability.  When deep in the 1D regime ($\sigma \ll 1$, as employed here) the roton instability arises for $g<0$ \cite{sinha_2007} [Fig. \ref{fig:stab}(b)]; however, for $\sigma \gappeq 1$ the roton instability  shifts to $g>0$, as predicted in Ref. \cite{giovanazzi_2004}.

\section{Dark soliton solutions}

Having established the stability of the homogeneous quasi-1D dipolar BEC, we move on to the dark soliton solutions themselves.  It is known that the purely local GPE is integrable and supports a family of dark soliton solutions for repulsive contact interactions ($g>0$) \cite{zakarov_1973,kivshar_1998}, 
\begin{equation}
\psi_{\rm s}(z,t)=\sqrt{n_0}\left[ \beta \tanh \frac{\beta (z-Z)}{\xi}+i \frac{v}{c} \right]e^{-i \mu t/\hbar}.
\label{eqn:ds}
\end{equation}
Here $\beta=\sqrt{1-v^2/c^2}$ and $Z(t)=z_0+vt$, where $z_0$ is the soliton's initial position and $v$ its speed.  Stationary solitons have a node of zero density and a phase slip of $\pi$, whereas a $v=c$ soliton has no phase or density contrast from the background fluid. The soliton energy decreases with increasing speed \cite{kivshar_1998}, leading to the analog of a particle with negative effective mass \cite{busch_2000}.  Meanwhile the density minimum of the soliton scales as $n_{\rm min}/n_0=\beta^2$ \cite{frantzeskakis_2010}.

Noting that dark solitons are stationary solutions in a moving frame, we numerically seek solutions to the GPE (\ref{eqn:gpe}) with a Galilean boost term $i \hbar v \partial_{z} \psi$ in the Hamiltonian.   Discretizing $\psi$ on a 1D spatial grid (spacing $dz=0.1\xi$) we minimise the discretized Hamiltonian with respect to changes in $\psi$ using a bi-conjugate method \cite{winiecki}. Starting with the non-dipolar dark soliton solution (\ref{eqn:ds}), this method leads to convergence to the required dipolar soliton solution, within the entire speed range $0\leq v \leq c$.  The numerical box (up to $\pm800\xi$) is large enough to mimic the infinite limit.  

The numerically-obtained dark solutions propagate with constant speed and permanent form (no dispersion or radiative losses) when simulated within the lab-frame GPE \cite{crank}, confirming their solitonic character.  Dark solitons require the mean-field potential of the homogeneous system to be net repulsive \cite{cuevas_2009,andreev_2014}, i.e. $\mu_0>0$, the same as the condition for phonon stability.

We illustrate the solutions through the $v=0$ soliton: Fig. \ref{fig:1sol} depicts its spatial density profile $n(z)$ as a function of $\varepsilon_{\rm dd}$ for the limiting angles (a) $\theta=0$ and (b) $\theta=\pi/2$.  The former is indicative of the general behaviour for  $\theta<\theta_{\rm m}$, and the latter for $\theta > \theta_{\rm m}$ (for $\theta=\theta_{\rm m}$ one recovers the non-dipolar soliton throughout).  The soliton size is characterised by the dipolar healing length $\xi$ (hence motivating this choice of units).  Away from the unstable band of $\varepsilon_{\rm dd}$, the profile approximates the $\tanh$-squared density profile of the non-dipolar soliton.  As the phonon instability is approached (for $g>0$) the profile 
diverges in width; this is related to a cancellation between the local interactions arising from the explicit van der Waals interactions and an implicit local contribution to the dipole-dipole interactions \cite{mulkerin_2013}.  Meanwhile, as the roton instability is approached, prominent density ripples form about the soliton core (see inset).   These finite-$k$ corrugations arise due to the mixing of the roton with the soliton state, with analogous ripples arising for vortices \cite{yi_2001,wilson_2008,mulkerin_2013}.   With increasing speed, the soliton depth decreases in line with the non-dipolar soliton, and the above qualitative behaviour remains, albeit with weaker ripples as the speed is increased.  The soliton phase profile is insensitive to $\varepsilon_{\rm dd}$ and $\theta$.

The soliton modifies the mean-field dipolar potential $\Phi_{\rm 1D}(z)$ in a non-local manner.  We can glean some insight into this by considering the behaviour at long range.  Expanding the dipolar pseudo-potential Eq. (\ref{eqn:U}) around infinity gives $U_{\text{1D}}(u)=U_0(4|u|^{-3}-24|u|^{-5}+\dots)$ \cite{abramowitz_1984}.  Then, taking the non-dipolar soliton (\ref{eqn:ds}) as an ansatz, the asymptotic form of $\Phi_{\rm 1D}$ follows as,
\begin{equation}\label{eqn:phi_a}
\Phi_{\rm 1D}(z)\simeq \Phi_0\left(1-\frac{\beta n_{0}\xi l_{\perp}^3}{e^{2\beta z_0/\xi}|z_0-z|^{3}}\right),
\end{equation}
where $z_0$ is a short-range cut-off to account for the asymptotic behaviour of the mean-field dipolar potential.
  The two terms represent the background and soliton contributions to $\Phi_{\rm 1D}$, respectively. The $1/z^3$ decay of $\Phi_{\rm 1D}$ demonstrates that the soliton appears as a localized giant dipole, a key result of our work.  This is intimately connected to the exponentially-fast decay of the soliton density profile; at large distances the soliton profile scales as $n(z) \simeq (n_0-n_{\rm min})e^{-z/\xi}$.  In contrast, a slow power-law-decaying density profile would render the object as an extended dipole at all scales, as for vortices in 2D \cite{mulkerin_2013}.  The negative sign on the soliton contribution suggests that the soliton may be viewed as a collection of {\it anti-dipoles} superposed on a homogeneous background of conventional dipoles.  This observation will help us below to interpret the modified form of the soliton-soliton collisions as due to the interaction of these anti-dipoles.

\section{Soliton-soliton collisions}

We can expect the non-locality of the solitons to modify their interactions, which we examine next through their collisions. For simplicity we take the collisions to be symmetric, i.e. same incoming speed for both solitons. The corresponding analytic two-soliton solution to the non-dipolar 1D GPE is \cite{frantzeskakis_2010},
\begin{equation}
\psi_{\rm 2s}(z,t)=\{F(z,t)/G(z,t)\}\exp(-i\mu t/\hbar)
\end{equation}
 with,
 \begin{eqnarray}
 F(z,t)&=&2\sqrt{n_0}\left[(1-2v^2/c^2)\text{cosh}(2v\beta t/\xi)\right. \nonumber
 \\
&-& \left. (v/c)\cosh(2\beta z/\xi)+i2(v/c)\beta \text{sinh}(2v\beta t/\xi )\right], \nonumber
\\
  G(z,t)&=&2\text{cosh}(2v\beta t/\xi)+2(v/c)\text{cosh}(2\beta z/\xi). \nonumber
\end{eqnarray}
Conventionally, solitons are known to repel during their collision with two regimes of dynamics: for low incoming speeds $v<0.5c$ the solitons reflect at short-range, with a sudden large repulsion akin to hard-sphere collisions, while for higher speeds $v\geq 0.5c$ the solitons transmit \cite{theocharis_2010,weller_2008}.  We focus on slow collisions $v=0.1c$, commenting on the wider speed dependence below.  We adopt fixed dipole parameters corresponding to $^{168}$Er \cite{aikawa_2012}, i.e. $\varepsilon_{\rm dd}=0.4$.  Figure 3 shows some example soliton collisions, simulated via the 1D dipolar GPE \cite{crank}, for different values of polarization angle $\theta$.  The solitons emerge from the collisions with unchanged profile and no radiative loss, further supporting their solitonic character.  At (b) the magic angle $\theta_{\rm m}$, one recovers the non-dipolar dynamics in which the solitons ``bounce'' at short-range \cite{theocharis_2010}.  For (a) $\theta<\theta_{\rm m}$, the soliton repulsion is significantly enhanced, turning them at increased separation, indicative of the role of the non-local interactions.  Here the dipoles are net attractive but the solitons themselves (which, recall, behave as anti-dipoles) are effectively net repulsive.  Vice versa, for (c) $\theta>\theta_{\rm m}$ the dipole-dipole interactions introduce an {\it attraction} between solitons; however, the effect of this is barely visible due to the dominance of the hard-sphere-like repulsion at short-range.

The non-local repulsive/attractive modifications to the soliton collisions become more pronounced with increasing $\varepsilon_{\rm dd}$ and $\sigma$ due to the increasing strength of the effective 1D dipole-dipole interactions.  Meanwhile, they become less pronounced with increasing soliton speeds; this is due to the reduced mean-field potential generated by faster and shallower solitons, as evident from the $\beta$ dependence in Eq. (\ref{eqn:phi_a}).  While the elastic collisions in Fig. 3 are representative of most of the stable $(\theta,\varepsilon_{\rm dd})$-space, collisions become inelastic close to the phonon instability or roton instability; then sound waves are shed during the collision, causing a slight (few percent) increase in their outgoing speeds.

\begin{figure}[t]
\centering
\includegraphics[width=0.99\columnwidth]{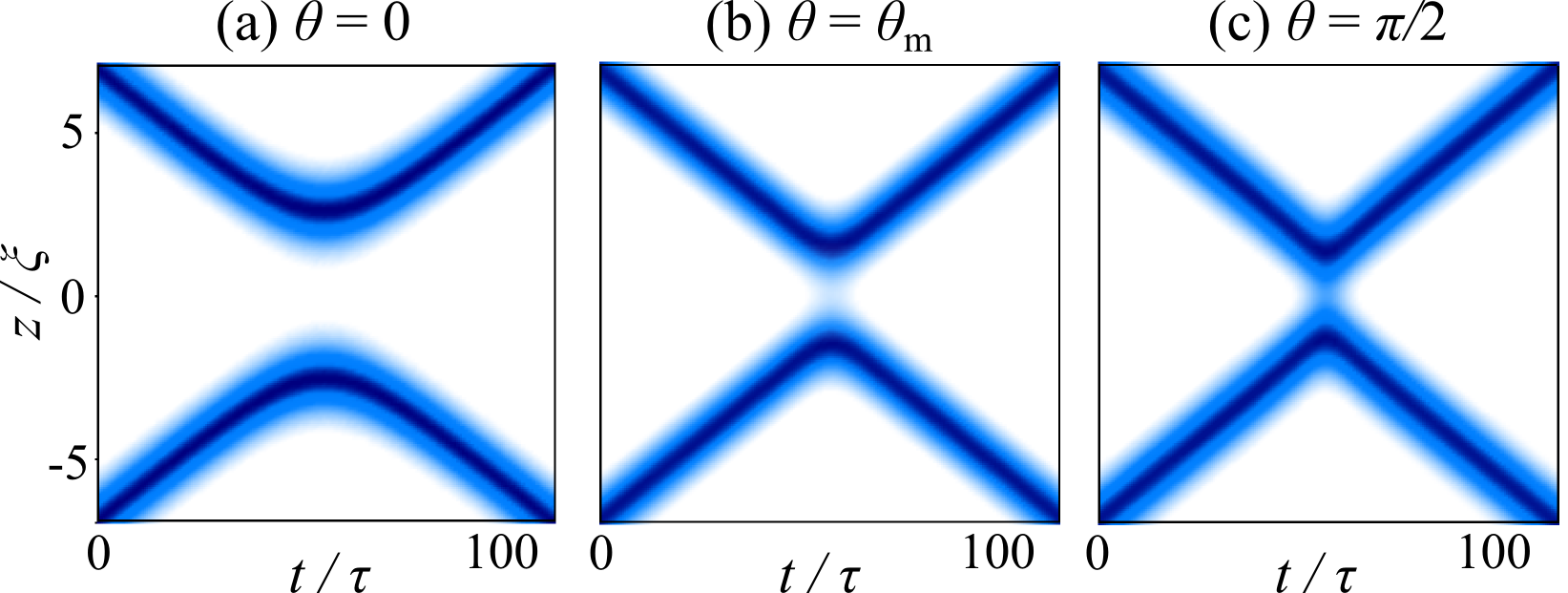}
\caption{(Color online) Collisions of dipolar dark solitons with incoming speeds $v=0.1c$ for $\varepsilon_{\rm dd}=0.4$ and polarization angles (a) $\theta=0$, (b) $\theta=\theta_{\rm m}$ and (c) $\theta=\pi/2$. 
}\label{fig:2sol}
\end{figure}

\section{Two-soliton interaction potential and bound states}

\begin{figure}[t]
\centering
\includegraphics[width=1\columnwidth]{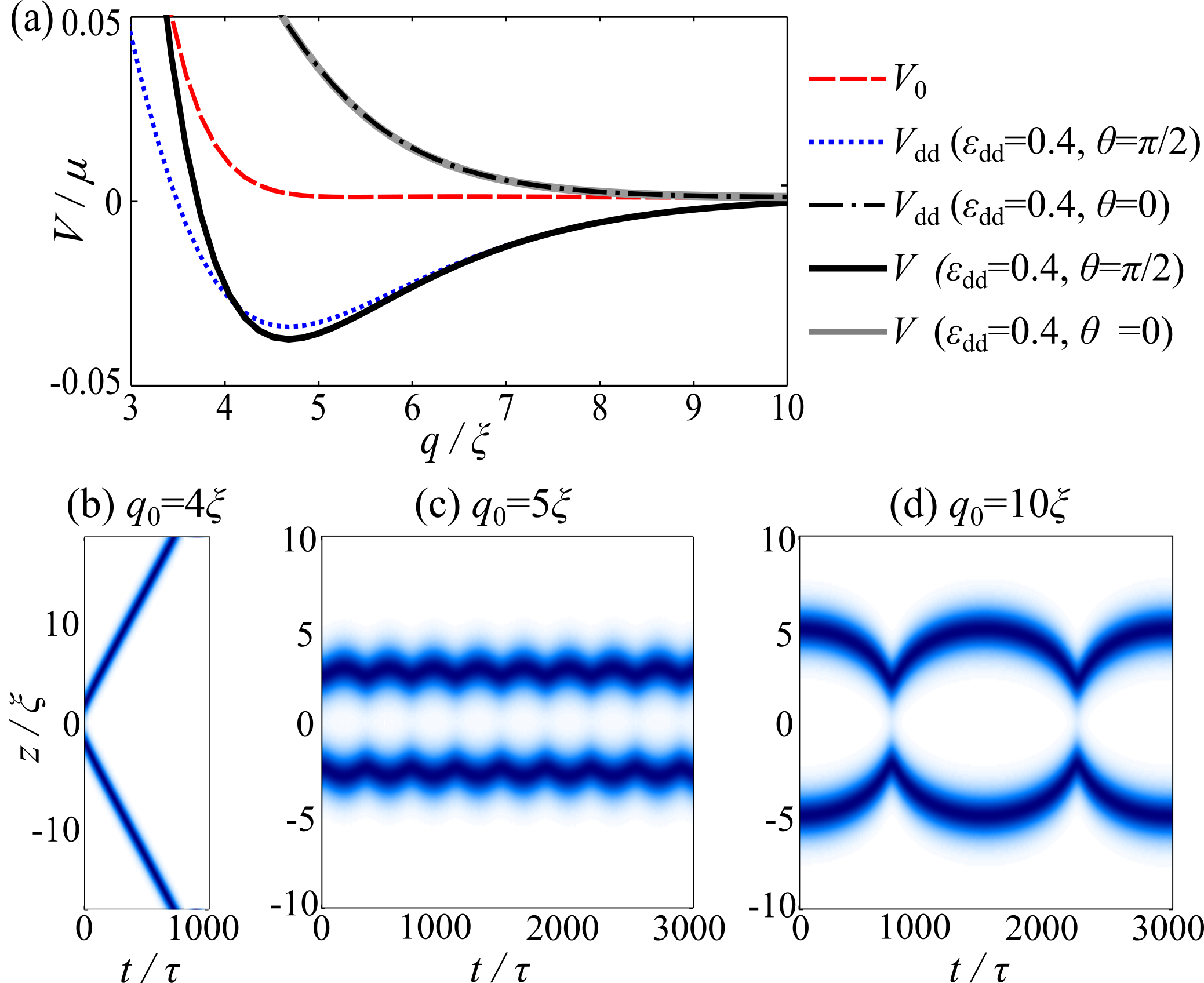}
\caption{(Color online) (a) Soliton-soliton interaction potential $V$ as a function of separation $q$  for $\varepsilon_{\rm dd}=0.4$ and $\theta=\pi/2$ (estimated using the non-dipolar two-soliton solution in the low speed limit with arbitrary value $v=0.001c$).  The local minimum indicates a bound state.  The non-dipolar $V_{\rm 0}$ and dipolar $V_{\rm dd}$ contributions are indicated. (b)-(d)  Corresponding GPE dynamics starting from two stationary solitons a distance $q_0$ apart.  For (a) $q_0=4\xi$, the solitons repel, while for (b) $q_0=5\xi$ and (c) $q_0=10\xi$, bound state oscillations are evident. }\label{fig:bound}
\end{figure}

We can further understand the soliton-soliton interaction through an effective particle-like interaction potential.   We calculate the soliton's energy, relative to the background, as,
\begin{equation}
E_{\rm tot}=E_{\rm 0}+E_{\rm dd},
\end{equation}
where,
\begin{equation}
E_{0}=\int \left[\frac{\hbar^2}{2m}|\partial_z \psi|^2+\frac{g}{2}(|\psi|^2-n_0)^2 \right]{\rm d}z,
\end{equation} 
is the non-dipolar energy (sum of kinetic and van der Waals interaction energies) and,
\begin{equation}
E_{\rm dd}=\int \frac{1}{2}\Phi_{\rm 1D}|\psi|^2~{\rm d}z,
\end{equation}
 is the dipolar contribution.   For two solitons at $z_1$ and $z_2$ (defined by their density minima), with separation $q=|z_1-z_2|$, we define the interaction energy as,
 \begin{equation}
 V(q)=E_{\rm tot}(z_1,z_2)-E_{\rm tot}(z_1)-E_{\rm tot}(z_2).
 \end{equation}  
 We estimate $V(q)$ semi-analytically based on the non-dipolar two-soliton solution with small incoming speed (for larger speeds, the  effective masses of the solitons change considerably during the collision, complicating this particle-like picture).  Figure \ref{fig:bound}(a) shows the soliton-soliton interaction potential, again for $^{168}$Er parameters.  The non-dipolar contribution $V_{\rm 0}$ (dashed red line) dominates only at short-range, consistent with the repulsive bouncing of two non-dipolar solitons [Fig.~3(b)]. 
 For $\theta=0$, the dipolar interaction potential $V_{\rm dd}$ (black dot-dashed line) is repulsive and non-local, consistent with the bouncing at increasing separation observed in Fig. 3(a).  However, for $\theta=\pi/2$, $V_{\rm dd}$ (blue dotted line) is attractive at long-range.  The playoff between this non-local attraction and the short-range repulsion from $V_{\rm 0}$, conspires to form a total potential (solid black line) analogous to the Lennard-Jones inter-atomic potential: an energy minimum at finite $q$, with rapidly increasing potential at short-range and slowly increasing potental at larger range.  This raises the prospect of forming a two-soliton bound state, analogous to a diatomic molecule.  
 
 The possibility of supporting a two-soliton bound state is probed through GPE simulations starting with two stationary solitons a distance $q_0$ apart, shown in Fig. \ref{fig:bound}(b)-(d). For sufficiently small $q_0$ (a), the solitons initially feel a strongly repulsive short-range interaction, repel and acquire sufficient kinetic energy to escape to infinity.  For greater separations, e.g. (b) and (c), the solitons initially have negative potential energy and are restricted to execute oscillations in $q$ about the potential minimum, analogous to vibrational modes of a diatomic molecule.  For large initial separations $q_0 \gg q_{\rm min}$, the solitons experience a weak gradient in $V(q)$ and undergo slow oscillations.  
 
Well away from the roton instability and phonon instability, these bound-state oscillations persist ad infinitum (within the zero-temperature GPE).  However, close to these instabilities, repeated sound emission during the collisions leads to the counter-intuitive situation where the solitons {\it increase} their oscillation amplitude and ultimately {\it escape} the bound state by {\it losing} energy; this is related to the negative effective mass of the soliton and is analogous to anti-damping effects in external potentials \cite{busch_2000}.   
 
 \begin{figure}[t]
\centering
\includegraphics[width=1\columnwidth]{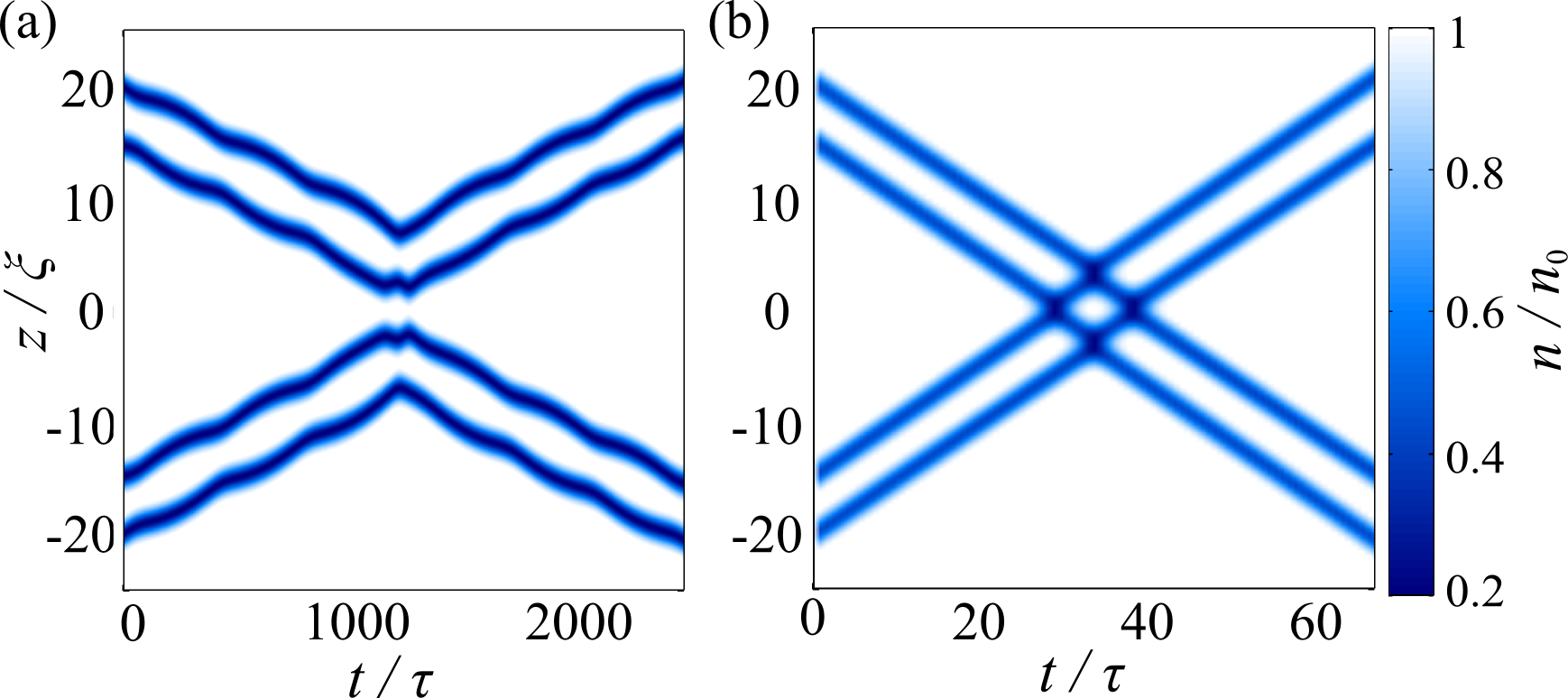}
\caption{(Color online) Collision of two counter-propagating bound states, with speeds (a) $v=0.01c$ and (b) $v=0.5c$  In both cases, the bound states emerge unscathed from the collision, with the only net effect being a phase shift of the outgoing waves (for the high speed case this phase shift is only just visible).}\label{fig:bound2}
\end{figure}
 
 We further investigate the properties of the bound state by considering its interaction with another bound state.  Figure \ref{fig:bound2} depicts the collisions of two counter-propagating bound states at (a) low and (b) high speed.  These moving bound states are formed by repeating the above method for forming bound states, but where both initial solitons have the same non-zero speed.  First, it is worth observing from the plot that the bound states are stable to centre-of-mass motion at constant speed. For both low and high speed, the two bound states emerge unscathed from the collision, with their original speeds and with no radiative losses.  The only net difference is that the outgoing bound states both feature a translational offset, termed a phase shift.   This is considerably larger for the low speed case, and is just visible in the high speed collision.  At low speed, the bound states appear to bounce off each other, just like in the collision of two slow dark solitons [Fig. 3], while at high speed the bound states appear to pass through each other, again analogous to the corresponding behaviour of two fast colliding dark solitons \cite{frantzeskakis_2010}.  Note that for the high speed case, the solitons in each bound state appear to move in parallel; this is simply because the period of the bound state oscillation is considerably longer than the timescale of the figure and the collision.  We have repeated these simulations over a wider rage of incoming speeds, and find the same qualitative soliton-like behaviour throughout. These results demonstrate the striking property that the bound states themselves behave like solitons in their interactions with other bound states.  
 
Furthermore, analysing the collision between a bound state and a single incident dark soliton shows the same behaviour, with the single soliton and bound state emerging unscathed from the collision, barring a phase shift.

\section{Conclusions}

In summary, we have studied self-trapped non-local dark solitons supported within quasi-1D dipolar BECs.  The solitons acquire modified profiles, including ripples associated with roton excitations.  The solitons approximate giant localized dipoles, and have non-local soliton-soliton interactions, controllable through the direction of polarization of the dipoles.  When attractive, and in combination with the conventional short-range repulsive interaction, unconventional dark soliton bound states can be realized.  These bound states are stable to centre-of-mass propagation at constant speed. Remarkably, they act themselves like solitons during collisions, emerging with unchanged form and speed.

Some of these effects are analogous to predictions for vortices in two-dimensional dipolar condensates \cite{mulkerin_2013}.  There, vortices bear similar ripples about the core, with these ripples being a common manifestation in dipolar condensates in the vicinity of a sharply-varying density profile.  However, the different density profiles of solitons and vortices lead to signficantly different results.  For a vortex, the profile scales asymptotically as $1/r^2$ back to the background density; this is sufficiently slowly varying that it renders the vortex as an {\it extended} dipole at all scales.  In contrast, the asymptotic part of the soliton density profile decays exponentially fast to the background, such that the soliton resembles a localized dipole at long range.  These differences, in turn, lead to different functional forms of their interaction potentials: for solitons the interaction potential scales with the dipolar form $1/r^3$ at long range, while for vortices the most slowly decaying terms go as $1/r^2$ and $\ln(r)/r^3$. For vortices this is in addition to the usual long-range, isotropic interaction arising from hydrodynamic effects.

The above soliton behaviour occurs robustly across a wide parameter space, accessible to dipolar BEC experiments \cite{griesmaier_2005,beaufils_2008,lu_2011,aikawa_2012} and with current dark soliton generation and imaging capabilities \cite{becker_2008,stellmer_2008,weller_2008}.  These dipolar dark solitons extend the range of physics of dark solitons as mesoscopic probes of quantum physics \cite{anglin_2008} to include the interplay with magnetism.  Moreover, the effectively instantaneous non-local soliton interaction, and its capacity to be experimentally tuned at will, offers intriguing possibilities for the controlled study of non-locality in complex networks \cite{rotschild_2006}, soliton gases \cite{el_2005} and super-solitons \cite{novoa_2008}.  

{\it Note Added:} After completing this research we became aware of the manuscript \cite{polish} which obtains results related to ours.

\begin{acknowledgements}
MJE and NGP acknowledge support by EPSRC (UK) Grant No. EP/M005127/1. TB acknowledges support from EPSRC (UK).  DO acknowledges support from NSERC (Canada).  
\end{acknowledgements}

\end{document}